\documentclass[superscriptaddress, nofootinbib, prd]{revtex4}[12pt]
\usepackage{amsmath}
\usepackage{bm}
\usepackage{euscript}
\usepackage[pdftex]{graphicx}
\usepackage{xspace}
\usepackage{xfrac}
\usepackage{enumerate}
\usepackage{braket}
\usepackage{float,fancyvrb}
\usepackage[T1]{fontenc}
\usepackage{lmodern}
\graphicspath{{./Figures/}}
\usepackage{wrapfig}
\usepackage{caption}
\usepackage{scrextend}
\usepackage{comment}

\usepackage{url}
\usepackage{amssymb}

\newcommand{\uu}[1]{\ensuremath{\, \mathrm{#1}}} 

\def\ea0{\mbox{ $ea_0$}}

\def\mF0{\mbox{ $\mu\Phi_0$}}
\def\mF0rtHz{\mbox{ $\mu\Phi_0/\sqrt{\rm Hz}$}}

\begin{document}
\title{Directional Detection of Dark Matter using Spectroscopy of Crystal Defects}


 \author{Surjeet Rajendran}
 \affiliation{Berkeley Center for Theoretical Physics, Department of Physics, University of California, Berkeley, CA 94720}
 
  \author{Nicholas Zobrist}
 \affiliation{Department of Physics, University of California, Santa Barbara CA 93106, USA}

 \author{Alexander O. Sushkov}
 \affiliation{Department of Physics, Boston University, Boston, Massachusetts 02215, USA}
 \affiliation{Photonics Center, Boston University, Boston, Massachusetts 02215, USA}

 \author{Ronald Walsworth}
 \affiliation{Department of Physics, Harvard University, and Harvard-Smithsonian Center for Astrophysics, Cambridge, MA 02138}
 
 \author{Mikhail Lukin}
 \affiliation{Department of Physics, Harvard University, Cambridge, MA 02138}

\date{\today}

\begin{abstract}
We propose  a method to identify the direction of an incident Weakly Interacting Massive Particle (WIMP) via induced nuclear recoil.  Our method is based on spectroscopic interrogation of quantum defects in macroscopic solid-state crystals . When a WIMP scatters in a crystal, the induced nuclear recoil  creates a tell-tale damage cluster, localized to within about 50 nm,  with an orientation to the damage trail that correlates  well with the direction of the recoil  and hence the incoming WIMP.  This damage cluster induces strain in the crystal, shifting the energy levels of nearby quantum defects. These level shifts can be measured optically (or through paramagnetic resonance) making it possible to detect the strain environment around the defect in a solid sample. As a specific example, we consider nitrogen vacancy centers in diamond, for which high defect densities and nanoscale localization of individual defects have been demonstrated. To localize the millimeter-scale region of a nuclear recoil within the crystal due to a potential dark matter event, we can use conventional WIMP detection techniques such as the collection of ionization/scintillation. Once an event is identified, the quantum defects in the vicinity of the event can be interrogated to map the strain environment, thus determining the direction of the recoil. In principle, this approach should be able to identify the recoil direction with an efficiency greater than 70 \% at a false positive rate less than 5\% for 10 keV recoil energies.  If successful, this method would allow for directional detection of WIMP-induced nuclear recoils at solid state densities, enabling probes of WIMP parameter space below the solar neutrino floor.  This technique could also potentially be applied to identify the direction of particles such as neutrons whose low scattering cross-section requires detectors with a large target mass. 
\end{abstract}

\maketitle
\section{Introduction} 
\label{sec:intro}
Weakly Interacting Massive Particle (WIMP) dark matter  is one of the most compelling dark matter (DM) candidates \cite{Gershtein:2013iqa}.   The weak scale is ultimately responsible for the origin of mass in the Standard Model and it is reasonable that it also sets the scales relevant for DM physics. Theories that attempt to explain the hierarchy problem naturally produce weak scale particles that interact  through processes mediated by the Higgs or electroweak gauge bosons. This is true not just in theories such as supersymmetry (which are presently heavily constrained by the LHC) but also in frameworks such as the relaxion where fermions at the weak scale carrying electroweak quantum numbers are a natural expectation \cite{Graham:2015cka}. Importantly, WIMPs have a calculable abundance -- thermal freeze-out of WIMPs naturally yields a cosmic abundance consistent with the observed DM density. 

Experimental work over the past three decades has cut deep into WIMP parameter space, with current experiments probing the possibility that DM may scatter via Standard Model interactions through the Higgs boson. These experiments are soon expected to hit a major background -- the coherent scattering of neutrinos from the Sun \cite{Strigari:2009bq}. WIMP DM experiments utilize a variety of handles to reject a number of radioactive backgrounds, such as the fact that  these radioactive backgrounds will typically scatter more than once in the detector, unlike the elastic scattering of DM. Unfortunately, the coherent elastic scattering of neutrinos from an atomic nucleus has the same event topology as DM scattering and the next generation of DM experiments are expected to be sensitive to solar neutrinos.  If this background cannot be rejected, WIMP detection would require statistical discrimination of a small WIMP signal over a large background. This implies that the sensitivity of the detectors would only scale as $\sqrt{V}$ where $V$ is the volume of the detector. Since WIMP detectors are already at $V \sim \text{m}^3$, continued progress would rapidly require prohibitively large detectors. 

One way to reject this background would be to identify the direction of the nuclear recoil induced by the collision of the DM (or neutrino) \cite{Nygren:2013nda, Battat:2016pap}. With such directional detection capability, one can make use of the fact that, due to momentum conservation, when a solar neutrino collides with a nucleus, the recoiling nucleus has to move away from the Sun. One could then reject all events that are pointed away from the known location of the Sun, eliminating the neutrino background. Incident WIMPs are expected to be isotropic; and thus by focusing only on events where the recoil is not along the direction of the Sun, one will be able to only look at events caused by DM. Such a directional detector will suffer a loss of sensitivity of $\sim$ 50 percent while dramatically reducing the neutrino background. 

It is thus of great interest to develop techniques to measure the direction of the nuclear recoil induced by a DM/neutrino collision. Not only would this permit continued exploration of WIMP parameter space below the solar neutrino floor where the WIMP can scatter via the Higgs boson \cite{Mayet:2016zxu}, but should the WIMP be discovered in the next generation of experiments, directional detection experiments would offer  a unique opportunity to measure the DM velocity profile. These measurements may even pave the way to discoveries of theorized galactic structures such as a dark disk or deviations from the naive Maxwell distribution of WIMP velocities. 

The technical problem that must be overcome for directional detection is the following. The scattering of DM/neutrino deposits energies $\sim 10 - 30$ keV. The direction of the induced nuclear recoil must be established in a detector with a large target mass, to overcome the tiny WIMP/neutrino cross-sections. To accommodate the large target mass without having to resort to enormous detector volumes, it is advantageous for the detector to be a high density material like a solid or a liquid. While there are excellent directional detection techniques in gas-based detectors \cite{Nygren:2013nda, Battat:2016pap}, there are no well established techniques for directional detection in high density materials. 

\begin{wrapfigure}{r}{4.5in}
\includegraphics[width=4.5in]{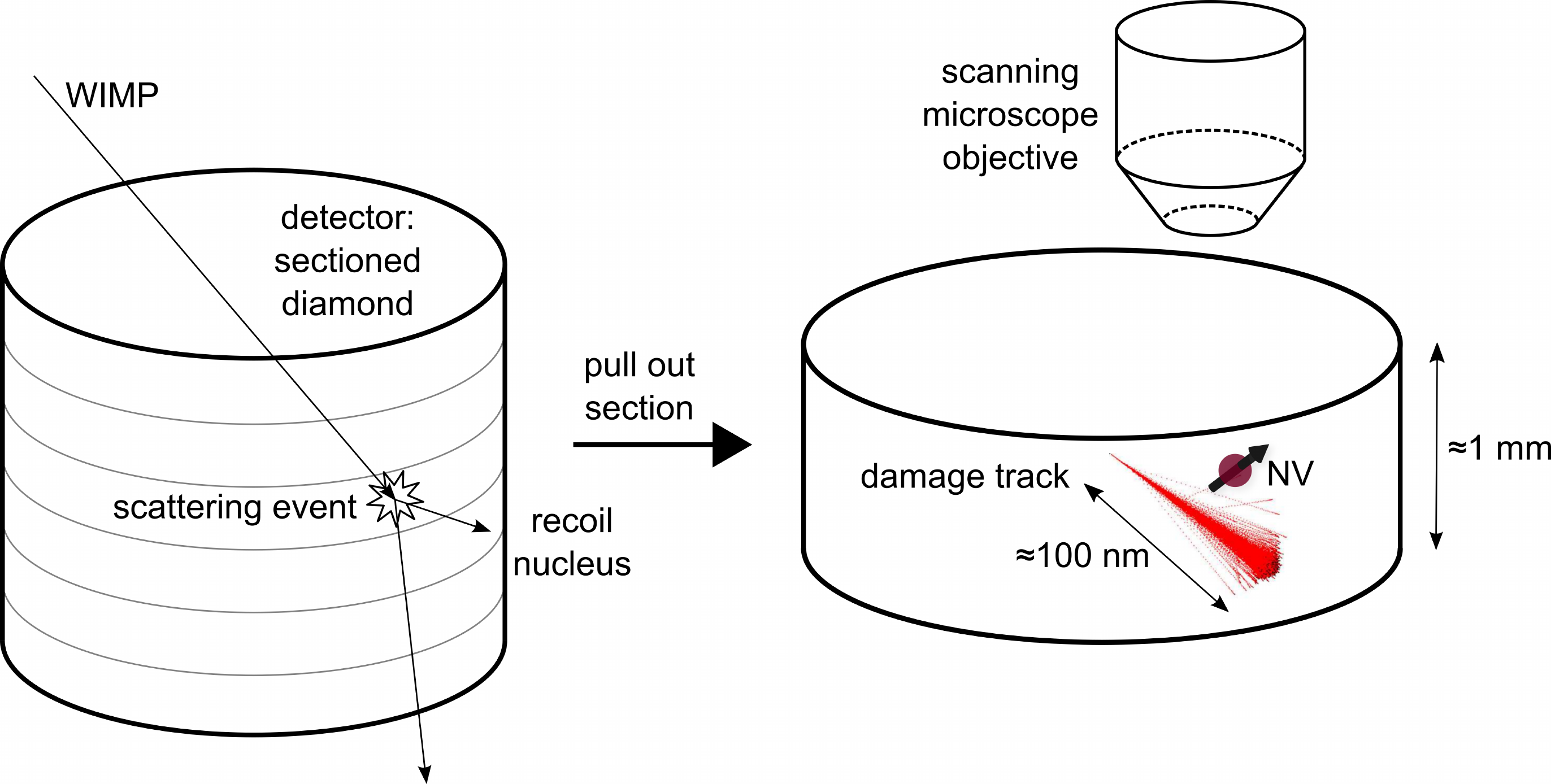}
\caption{
Left: Event identified by conventional methods. Right: Section of interest separately studied by superresolution methods.
}
\label{fig:SectionedDiamond}
\end{wrapfigure}

Here we propose a new directional detection scheme that can operate at solid state densities, with the ability to accommodate large target masses so that the concept could potentially be applied for DM detection. The basic idea employs crystals with point quantum defects.
When a WIMP scatters near one of these defects, the induced nuclear recoil creates a tell-tale damage cluster, localized to within $\sim 50$ nm, and with an orientation that  correlates well with the direction of the recoil. This damage cluster induces strain in the crystal and this strain shifts the energy levels of the nearby defects. These energy level shifts can be measured by exciting optical transitions, or ground state magnetic resonance spin-flip transitions, making it possible to map the strain environment in the detector crystal with spatial resolution on the order of the defect spacing. To identify the location of a nuclear recoil to a millimeter-scale volume, we can first use  conventional WIMP detection approaches such as the collection of ionization/scintillation radiation to identify  and localize potential dark matter events \cite{Agnese:2013rvf, Aprile:2016swn, McKinsey:2016xhn} within the crystal. Once an event is identified and localized, the defects in the vicinity of the event can be interrogated to determine the strain environment, thus identifying the direction of the recoil. See Figure \ref{fig:SectionedDiamond}.  If successful, this concept would open a new path to continue the probe of the theoretically well motivated WIMP. The phenomenology of quantum defects described above is found in a number of systems such as nitrogen vacancy (NV) centers  \cite{Jelezko,Childress2014}  and/or silicon vacancy (SiV) centers~\cite{Hepp2014} in diamond, paramagnetic F-centers in metal halides \cite{MetalHalide} or defects in Silicon Carbide \cite{Koehl}. 

Our present discussion is focused on NV centers in diamond, motivated by the fact that these defects are well studied,  with a number of experimental parameters and interrogation techniques well established. However,  technical considerations may make systems such as SiV centers in diamond or divacancies in SiC better suited for a DM search than NV centers. Specifically, the optical transition in SiV centers in diamond has a narrow linewidth at cryogenic temperatures ($\approx 5$~GHz at 77~K and $\approx 300$~MHz at 10~K~\cite{Evans2016}); and its frequency is both sensitive to strain~\cite{Sternschulte1994} and first-order insensitive to electric field. From a materials perspective, silicon carbide may be easier to work with than diamond, since it is commercially available as high-purity single-crystal wafers up to several inches in diameter, and divacancy defects have electronic level structure and coherence times that are very similar to NV centers~\cite{Christle2017}. If the proposed concept proves fruitful, we believe there will be strong motivation to study a wider class of crystal defects to identify optimal systems for specific applications. 

The paper is organized as follows. In section \ref{sec:damage}, we discuss the localized damage caused by DM scattering. In section \ref{sec:concept} we present a conceptual overview of how optically detected magnetic resonance (ODMR) of NV centers in diamond can be used to detect the resulting nanoscale strain patterns. In section \ref{sec:strain}, we discuss the technical details of this measurement and evaluate the efficiency of our measurement protocol. 

\section{Crystal Damage}
\label{sec:damage}
\begin{wrapfigure}{r}{0.5\textwidth}
\includegraphics[width=0.5\textwidth]{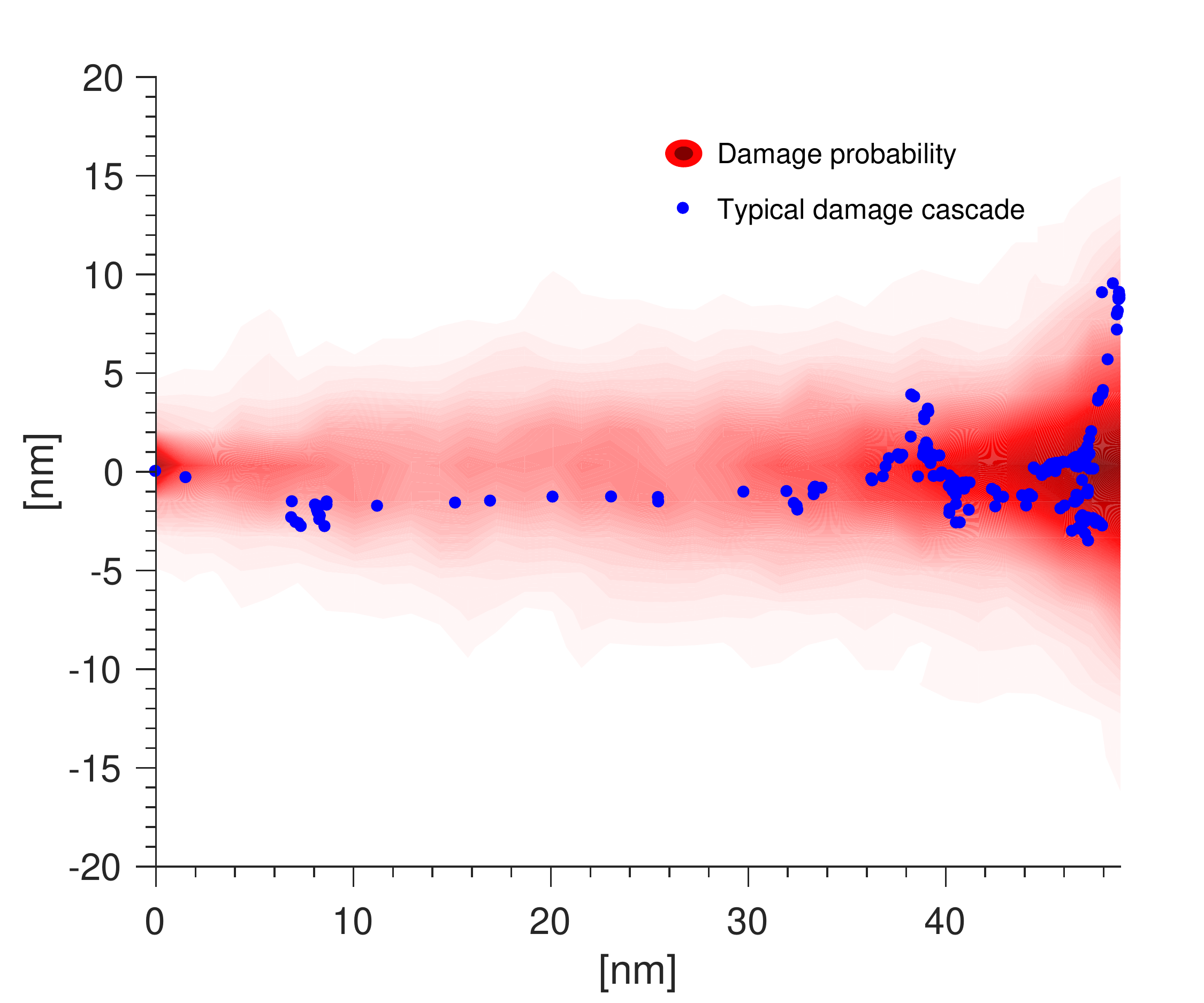}
\caption{
A typical damage cluster shape from a 30 keV ion injected into a crystal from a TRIM simulation. The cluster is well correlated with the direction of the injected ion. The ion should be thought of as the recoiling nucleus, coming from a WIMP scattering event.
}
\label{fig:Damage}
\end{wrapfigure}

The elastic collision of a conventional WIMP (mass $\sim 10$ GeV) with a nucleus  is expected to deposit energies $\sim 10 - 30$ keV. This energy is significantly larger than the lattice potential $\sim 10$ eV of typical crystals. The recoiling nucleus scatters with the lattice, creating a localized damage cluster consisting of interstitials and vacancies. The damage cluster created by such a collision can be modeled using a  Transport of Ions in Matter (TRIM) simulation  \cite{SRIM}. We consider a carbon lattice (appropriate for diamond) and input an initial carbon ion with energy $\sim 10 - 30$ keV. This initial ion represents the recoiling nucleus. The TRIM simulation captures the effect of this input ion and the result from a typical event is  shown in Figure \ref{fig:Damage}. These results indicate that one generically obtains about $\mathcal{O}\left(100  - 300\right)$ interstitial/defects created by the scattering -- this is consistent with the fact that the energy deposits are $\sim 10 - 30$ keV, with the lattice potential $\sim$ 10 eV. Futher, the damage trail is well correlated with the initial direction of the input ion. The damage is also asymmetric -- we typically see a larger number of dislocations at the end of the damage trail than the beginning. The damage trail itself is localized within $\sim$ 50 nm. Since the shape of this trail is well correlated with the recoil direction, its spatial characterization will lead to directional resolution  of the nuclear recoil and hence the incoming WIMP. It should be kept in mind that once the cluster is created, it is not easily destroyed since the typical barriers associated with defect migration are $\sim$ eV, implying thermodynamic stability at operating temperatures (300~K). 

This crystal damage creates strain in the lattice. To calculate the induced strain, we follow \cite{Strain}: the strain from a single vacancy/interstitial falls off as $1/r^3$, where $r$ is the distance from the defect. The total strain from the damage cluster can be calculated by adding the strain from each individual defect. Therefore the strain $\frac{\Delta x}{x}$ at a location 30 nm away from a single vacancy is $\frac{\Delta x}{x}\approx \left(\frac{0.3\text{ nm}}{30\text{ nm}}\right)^3\approx 10^{-6}$. This corresponds to a stress $P = Y\frac{\Delta x}{x}\approx 10^{6}\text{ Pa}$ where $Y\approx 10^{12}\text{ Pa}$ is the Young's modulus of diamond. This stress can be detected using the shift of the zero-magnetic-field transition frequency between the ground-state magnetic sub-levels of an NV center. The stress coupling coefficient~\cite{Togan2011} is $0.03\text{Hz/Pa}$, so the transition frequency shift is $\Delta f \approx 30\text{ kHz}$. This can be detected via NV ODMR using a standard ``clock'' measurement protocol, insensitive to magnetic fields, and should be compared to the the NV transition linewidth, which is limited by $1/T_1\approx 300\text{Hz}$ at room temperature~\cite{Kucscko}. We note that the spin-1 ground state of the NV center allows determination of the transverse and perpendicular components of the strain relative to the NV axis.

In order to reconstruct the shape of the crystal damage distribution, and thus the direction of the momentum of the scattered WIMP, the diamond should have NV center density of approximately $1/(30\text{ nm})^3$ (which has been demonstrated in bulk diamond samples \cite{Bassett1, Dima1}). Superresolution imaging  techniques \cite{Jaskula, Arai} can then be used to perform zero-field transition frequency  measurements of several individual NV centers near the scattering event location to map the strain profile within the diamond with few nanometer precision. Such a measurement method will provide good sensitivity to crystal damage, as it leverages the remarkable spin properties of NV centers (long spin lifetime), as well as the stiffness of diamond (large Young's modulus). Note that the density of interstitials/vacancies created by the WIMP scattering is locally a factor of $\sim 100$ larger than the NV center density -- thus there is a distinct and large local signal for crystal damage, such that the key technical challenge will be identifying the $\text{mm}^3$ region within the crystal where the WIMP scattering event occurred.

Before further describing a detector concept that can measure this damage, let us discuss the correlation between the damage cluster and the recoil direction. This correlation was estimated using the TRIM simulation as discussed above. We find that in nearly 70 percent of events with energy depositions $\sim$ 10 keV (or larger), there is a clear asymmetry ($1:2$) in the number of vacancies/interstitials in the beginning versus the end of the damage cluster, enabling head/tail discrimination of the recoil. The false positive rate is estimated to be less than 5 percent. Thus, even with a few events, it is possible to make a statistically robust discovery of DM. These asymmetries degrade significantly below $\sim$ 1  keV recoils (see Figure \ref{fig:distributions}), making this method most useful for identifying the direction of conventional WIMP (mass $\gtrapprox$ 1 GeV) DM. 

\section{Concept}
\label{sec:concept}

The results of the TRIM simulation suggest that with a NV center density of $1/\text{(30 nm)}^3$, the $\mathcal{O}\left(100\right)$ crystal defects created by WIMP-induced nuclear recoil within $\sim 50$ nm produce NV zero-field frequency shifts $\sim 10$ kHz, significantly larger than the NV transition linewidth $\sim$ kHz. Thus, by mapping the strain-induced NV zero-field frequency shifts around the damage cluster using superresolution techniques, the incident WIMP direction can be determined. But, how do we find the correct group of NV centers  in proximity to the recoil-induced damage? For a practical rate of WIMP scattering events, the crystal will need to have a large volume. 

The following protocol should allow coarse localization  ($\sim \text{mm}^3$) of the WIMP scattering event, making it realistic to identify the NV centers proximal to the damage trail, using a scaled-up version of existing NV ODMR technology. We consider a sectioned detector as shown in Figure \ref{fig:SectionedDiamond}, with each section of thickness $\sim$ mm (the lateral dimensions can be much larger, potentially $\sim$ m). Assume there is some collision in this detector -- this collision may be due to a WIMP, neutrino or radioactive background. We propose using standard techniques from conventional WIMP detectors to identify these events: for example, through the collection of scintillation/ionization we can identify the small number of single scattering events that could potentially be due to DM/neutrino scatterings. For WIMP scattering cross-sections of interest, we expect  $\mathcal{O}\left(10\right)$ events within the meter-scale detection volume over the course of about one year, whose direction would then have to be determined. Relying upon conventional WIMP detection techniques, the spatial localization of each of these events can potentially be known to a volume $\sim$ mm$^3$ \cite{McKinsey:2016xhn}.
Scintillation in diamond has been observed~\cite{Miller1966}, but its properties need to be investigated in detail in order to evaluate the feasibility of millimeter-scale event localization. Ionization (electron-hole production) in diamond, on the other hand, is very well studied, and is used in a variety of diamond-based radiation and particle detectors~\cite{Tapper2000}. In order to provide the necessary spatial resolution, charge extraction would need to be done with pixelated electrodes~\cite{Rebai2015}, which would likely introduce additional radiation backgrounds (see below), but this can potentially be controlled by careful fabrication, characterization, and discrimination~\cite{CDMS2012}.
The technical challenge then  reduces to the identification of the correct set of NV centers within each target $\sim \text{mm}^3$ volume of interest. Since  the crystal damage is stable, we can take significant time (several days) to study each region of interest to identify the direction of the recoiling nucleus. 


Once we identify the events of interest, the associated mm section of the detector is pulled out for further study. The $\text{mm}^3$ region of interest is interrogated via ODMR to identify the group of NV centers whose zero-field frequencies are significantly shifted due to damage-induced strain. This method is of course diffraction limited to a resolution $\sim \mu \text{m}$, the wavelength of the light. Superresolution optical imaging \cite{Jaskula} and/or strong magnetic field gradients \cite{Arai} can now be applied to this region for nano-scale resolution and ODMR of individual NV centers.  For example, with an applied external magnetic field gradient of $\sim 1 \text{ Tesla/cm}$ \cite{Arai} the induced frequency shift in NV centers separated by $\sim 30$ nm is $\sim 10$ kHz, larger than the linewidth of the NV center, permitting such resolution. This measurement protocol explains the need for sectioning the detector: resolution of the NV centers below the wavelength of light requires the application of large external magnetic field gradients or precisely shaped, intense optical fields \cite{Jaskula} that are most easily accomplished when the thickness of the section is not too big. Note that SiV centers in diamond are a promising alternative quantum defect, as they have an optical transition frequency that is sensitive to local crystal strain~\cite{Sternschulte1994}; and could provide nanoscale mapping of local strain in an optical-background-free manner, if the sample is cooled to cryogenic temperature where the strained SiV centers could be spectrally resolved without the need for superresolution imaging.

\section{Nanoscale Probe of Crystal Strain}
\label{sec:strain}
As outlined above, once the scattering events of interest are identified and localized to within $\sim$ $\text{mm}^3$ by conventional WIMP detection methods, the plate (of thickness $\sim$ mm) in which the event occurred is pulled out and examined. The crystal damage caused by the event is then probed at the nanoscale by mapping the resulting strain on nearby NV centers or other quantum defects. 

\subsection{Measurement Process}

As estimated above, a single vacancy or interstitial creates a stress of $\sim$ $10^6$~Pa at a location 30~nm away; and if there is an NV center at that location, the resulting shift of its ground state zero-field splitting is $\sim$ 30~kHz. A scattering event produces a damage track with length scale $\sim 50$~nm, containing 100-300 such vacancies. Our task is to localize and characterize this damage track, specifically extracting the track asymmetry and therefore initial recoil direction, starting from an initial localization accuracy of $\sim 1$~mm$^3$. 

For example, the sensitivity to local stress of a single optically resolved NV center can be estimated \cite{Kucscko} using $\eta = \frac{1}{C d\Delta/dP} \frac{1}{\sqrt{\tau_{\text{coh}} t}} \sim 100 \text{Pa}/\sqrt{\text{Hz}}$. Here, $C\approx 10^{-2}$ is the factor that accounts for initialization and readout imperfections, $\tau_{\text{coh}} \approx 1 \text{ ms}$ is the NV center coherence time (under the magnetic field-insensitive clock sequence), $d\Delta/dP \sim 0.03 \text{ Hz/Pa}$ is the stress coupling coefficient, and $t$ is the averaging time. 

We begin the event localization with wide-angle imaging of the 1~mm$^3$ volume of the detector material, using a CCD or CMOS camera. Each pixel on the camera images $\sim 1$~$\mu$m$^2$ area of the detector. Standard optical techniques can be adapted for imaging point sources within a high index of refraction crystal (e.g., diamond) with depth of field of about 1 micron. By focusing the excitation laser, we divide the detector into $\sim 1$~$\mu$m$^3$ voxels, each containing $\sim 3\times 10^4$ NV centers. If one such voxel contains the damage track, then several NV centers within it will exibit $\sim$MPa stress, which is detectable as zero-field frequency shifts with good signal-to-noise after 100 seconds of averaging. Since we are only imaging a 1~$\mu$m-thick section of the detector at a time, we have to repeat the imaging 1000 times in order to scan the entire 1~mm$^3$ volume, which takes a total of $\sim 10^5$~seconds -- about a day. Having thus localized the damage to a $1$~$\mu$m$^3$ voxel, we then employ optical superresolution techniques and/or strong magnetic field gradients, in order to extract the track asymmery.  Such nanoscale imaging will require extra overhead compared to diffraction-limited optical imaging; assuming measurement time of 10 seconds per NV center, this will take about 3 days. Therefore the entire damage track can be characterized on the time scale of several days. 


\subsection{Estimated Efficiency}
To reject the solar neutrino background, we must determine the initial nuclear recoil direction from the zero-field frequency shifts of the NV centers closest to the damage site. This pattern recognition problem may be approached in many ways, with the effectiveness of a particular method described by its efficiency: the percentage of events whose initial recoil direction is accurately inferred from the damage left in the crystal, for a given false positive rate. 

\begin{figure} 
\includegraphics[width=.9\textwidth]{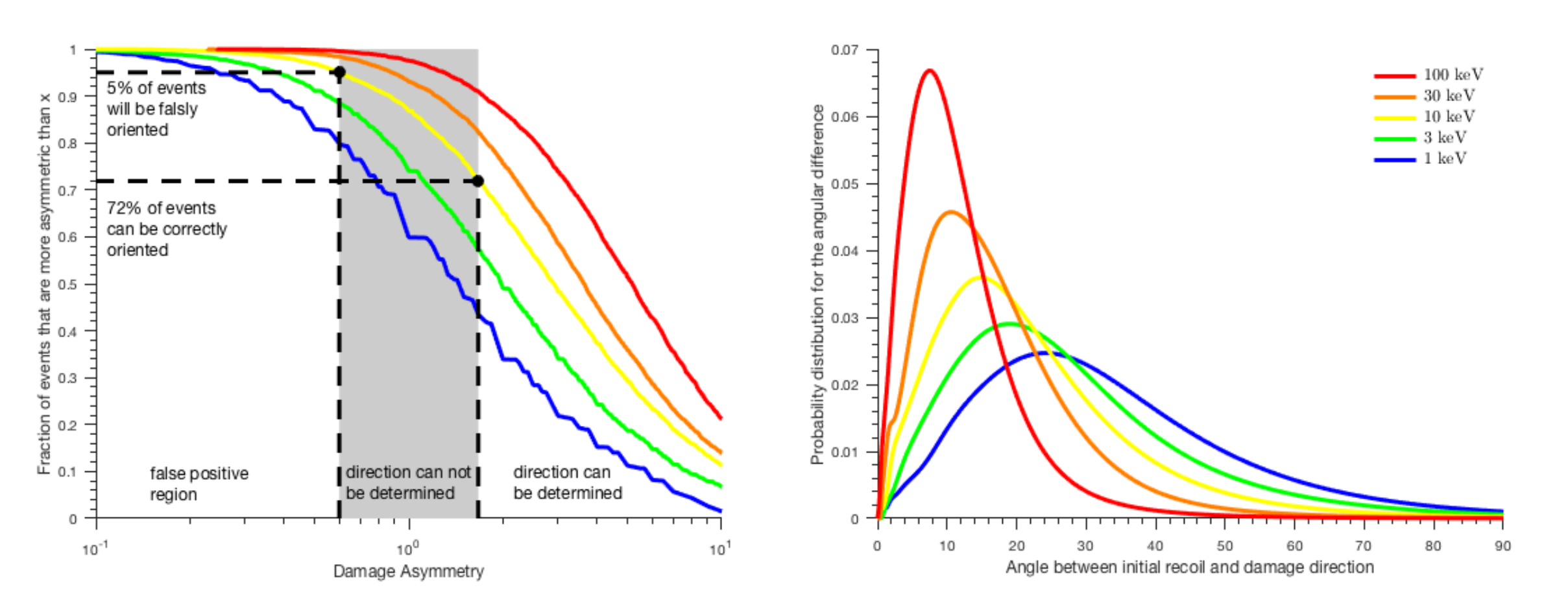}
\caption{ TRIM simulation results in diamond are plotted for different initial recoil energies. Left: the distribution of damage asymmetries is shown. An example asymmetry cut is shown with a 5\% false positive rate resulting in an efficiency of $\sim$ 70\%. Right: the angular difference between the initial recoil direction and the overall damage direction is shown. The two directions are more correlated for higher initial recoil energies.}

\label{fig:distributions}
\end{figure}

To find the maximum efficiency we must consider that a minority of damage tracks are asymmetric in a direction opposite to the initial recoil (asymmetry<1). Here, we define asymmetry as the ratio of the number of lattice interstitials and vacancies in the end third of the damage to the beginning third. The left side of Figure~\ref{fig:distributions} shows the distributions of asymmetry for different initial recoil energies in diamond, computed from the TRIM simulations discussed in section~\ref{sec:damage}. By cutting events with small asymmetry from consideration, we also remove the majority of events asymmetric in the wrong direction. These results give a maximum efficiency of $\sim$ 70\% and  $\sim$ 90\% at a 5\% false positive rate for 10 and 30 keV collisions in diamond, respectively.  

We tested this procedure by computing strain-induced zero-field frequency shifts for randomly placed NV centers for many different damage trails. We then used a simple analysis to infer the initial recoil direction for each damage trail from the NV strain map. At 30 keV our code achieved an efficiency of $\sim$ 50\% with an NV center density of $\sfrac{1}{\left(10 \text{ nm}\right)^3}$. We expect that use of a state-of-the-art pattern recognition algorithm will greatly increase the efficiency.

After cutting out events with low asymmetry, we reject the solar neutrino background by removing events that have an initial recoil direction pointing away from the Sun. However, not all damage directions are well-correlated with the initial recoil direction. The right side of Figure~\ref{fig:distributions} shows this relationship for different recoil energies. For lower energy events, slightly more than half the data will be cut.

\subsection{Preparation and Backgrounds}

Crystal defects (such as the NV center) are manufactured during chemical vapor deposition (CVD) and/or through irradiation followed by subsequent annealing of the crystal \cite{Bassett1, Dima1}, where defect densities as high as $1/(\text{5 nm})^3$ have been demonstrated. One may worry that the process of creation of this large density of NV centers might introduce additional backgrounds. For example, electron irradiation to create vacancies must deliver energies $\sim 20$ eV to displace a nucleus, which can be accomplished with electrons of energy $\sim$ MeV.  These electrons could be captured by the nuclei in the crystal (carbon/nitrogen),  render them unstable and cause additional radioactive background. The existence of this background would depend upon the choice of crystal and quantum defects used as strain sensors: for concreteness, we consider diamond with NV centers. First, electron capture proceeds through the weak interaction, leading to a capture cross-section $\sim 10^{-43} \text{ cm}^2$ for $\sim$ MeV electrons. Thus, to create a NV center density  $1/(\text{30 nm})^3$ in a meter-scale sample, we expect $\sim 1000$ electron capture events. These capture events can, for example, convert Nitrogen 14 to Carbon 14, which can then subsequently undergo radioactive decay. Considering the life-time of Carbon-14 $\sim$ 6000 years, we expect $\sim \mathcal{O}\left(1\right)$ events per year. Of course, these decays by themselves are not a background for our WIMP directional detection scheme: Carbon 14 decays through beta emission; hence the resulting electron track would  distinguish these events from those of WIMP-induced nuclear recoil and strain. A problem would arise only if there were an enormous number of such decays, which would lead to a large radioactive background that produces a large number of crystal defects. But, since the expected number of electron captures is low ($\sim 1000$ in a $\text{m}^3$ sample), the production process is not likely to recreate a major background. It should be noted that $\sim$ MeV electrons are not likely to excite unstable nuclear levels in these systems -- but even if they are produced, as long as those states decay before the annealing time  they will not be a problem.

Crystal damage caused by other radioactive sources is a potential cause for concern. If the event rate from these backgrounds is large enough to cause crystal damage within the initial localized region ($\sim \text{mm}^3$), we will not be able to distinguish the damage caused by radioactivity from DM/neutrino scattering. Note that this problem arises only for nuclear recoils -- the damage trail caused by electron recoils where the electron slowly loses energy $\sim 10 \text{ eV}$/angstrom over a long distance is significantly different from the DM/neutrino induced nuclear recoil where the deposited energy is lost within $\sim$ 30 nm. Since the lattice potentials for a typical crystal are $\sim 10 - 30$ eV, the damage trail from electron recoils may cause defects with a density $\mathcal{O}\left(\text{few}\right)/\left(\text{30 nm}\right)^3$ spread over a longer distance than the DM/neutrino induced nuclear recoils that lead to defect densities $\sim \mathcal{O}\left(100 - 300\right))/\left(\text{30 nm}\right)^3$. For nuclear recoil backgrounds to be insignificant, we would need less than one damage trail within the initial localized volume $\sim \text{mm}^3$, implying that the overall total nuclear recoil background be less than $\sim 10^{9}$ events/year  in a detector of volume $\sim \text{m}^3$. Background rates smaller than this have been achieved in bulk volume detectors \cite{Aprile:2015uzo}, especially in the inner parts of the detector where self-shielding effects are important. It is also important to note that the effects of this radioactive background quickly become sub-dominant with improvements in the initial localization of the event. 

Another factor to consider is the inherent lattice strain invariably present in the detector crystal. Such inherent strain is known to be present in diamond crystals, where it gives rise to inhomogeneous broadening of the NV $^3$A$_1$ -> $^3$E$_1$ electric dipole optical transition zero-phonon line~\cite{Chu2014}. For the typical inhomoheneous optical transition linewidth $\sim$GHz, we can estimate the lattice strain $\sim$MPa, which is on the same order as the expected strain caused by the target WIMP-induced nuclear recoil damage. However, it is possible to reduce this inherent lattice strain in diamond by a high-temperature anneal preparation process, which results in a two order-of-magnitude reduction of the inhomonegenous linewidth~\cite{Chu2014}. 
Additional discrimination is provided by the fact that the damage track due to a WIMP scattering event is very local  ($\sim50$~nm), whereas the inherent strain is likely to have a much larger length scale, even in polycrystalline diamond~\cite{Trusheim2016}. 
Careful detector preparation and strain spatial discrimination may allow us to use polycrystalline diamond with grain sizes on the order of hundreds of microns to millimeters, which would significantly simplify detector fabrication.

\section{Conclusion}
\label{sec:conclusions}

Directional detection via crystal defects offers a promising approach to probe WIMP dark matter with mass below $\sim 30$ GeV where the dominant neutrino background arises from the solar neutrino flux \cite{Mayet:2016zxu}. For the range of energies deposited by the scattering of such WIMPs, there is significant localized crystal damage that correlates well with the direction of the recoiling nucleus. This crystal damage is thermodynamically stable for long periods of time. Since we will at most have to resolve the direction of a handful of neutrino/WIMP events (utilizing conventional WIMP detection techniques to veto the other, far more dominant backgrounds), it is feasible to expend significant time (few days per event) to map this localized damage. As discussed above, the strain induced by WIMP-induced crystal damage should be measurable using nanoscale sensors such as NV centers in diamond, where we leverage the fact that locally (within 50 nm) the expected damage from dark matter scattering is $\mathcal{O}\left(100-200\right)$ times the typical mean density of lattice vacancies in diamond. 

The efficiency of our technique depends on the energy deposited in the scattering event. Presently, it is of most use for WIMPs with masses $\gtrapprox$ 1 GeV. The direct detection of lighter WIMPs, where the solar neutrino background is bigger, is an active area of research \cite{Bunting:2017net, Derenzo:2016fse, Essig:2016crl, Essig:2015cda, Graham:2012su,Essig:2011nj, Guo:2013dt} and it would be interesting to investigate the efficacy of the crystal defect approach within these emerging detection schemes. In particular, since the strain from localized crystal damage drops off as $\sim 1/r^3$, a larger density of NV centers might be able to resolve the weaker damage caused by light WIMPs. But, we would then have to identify this damage cluster by probing a larger number of NV centers, significantly increasing the time necessary to perform the measurement. This time could be reduced if the position resolution of the detection scheme that identifies the putative WIMP/neutrino events was improved. In our study, we assumed a position resolution $\sim$ mm$^3$. The number of NV centers that have to be interrogated to resolve the damage track scales linearly with this resolution volume. Thus, improvements in the position resolution will drastically cut down on the time necessary to identify the damage track, potentially permitting the use of higher NV center densities to study lighter WIMPs. For heavier WIMPs, where  isotropic backgrounds dominate, directional detection would still be interesting since one could try to leverage the annual modulation of the directional signal from the motion of the solar system in relation to the galactic dark matter flow. 

In addition to dark matter detection, these nanoscale sensors may also be of use in improving the angular resolution of bulk detectors necessary to detect particles with low cross-section, for example, neutrons. The kinematics of neutron scattering is similar to that of WIMP-nucleon interactions, and thus similar crystal damage ought to exist.  The study of such damage tracks could be used to calibrate future dark matter detectors and may also be of use in the context of non-proliferation. While NV centers in diamond are experimentally well studied, the phenomenology of crystal damage that can subsequently be probed by a nanoscale sensor is valid for a number of other systems. It would be interesting to explore a variety of such defects and identify a broader class of materials that might be more optimized for detecting a variety of other particles, including dark matter. 

\acknowledgements

We would like to thank the Heising-Simons Foundation for organizing a workshop at UC Berkeley where this idea was conceived; and also the Moore Foundation for organizing a meeting where the idea was further explored. We also thank Dmitry Budker, Daniel McKinsey, Matthew Pyle, Alp Sipahigil, Ruffin Evans, and Meesala Srujan for helpful conversations.  SR was supported in part by the NSF under grants PHY-1417295 and PHY-1507160,the Alfred P. Sloan Foundation grant FG-2016-6193, the Simons Foundation Award 378243 and  the Heising-Simons Foundation grant 2015-038. AOS was supported in part by the Heising-Simons Foundation grant 2015-039 and the Alfred P. Sloan Foundation grant FG-2016-6728.

{\bf Note Added:} While this work was  in preparation, \cite{Tommer} appeared. While both ideas consider the use of crystal defects for dark matter detection, our idea differs significantly from  \cite{Tommer}. In  \cite{Tommer}, the detection of the small number of crystal defects produced by the  scattering of light dark matter is proposed as a way of searching for sub GeV mass dark matter. Directional information is inferred through anisotropies in the creation of such defects.  Our focus is on the use of existing crystal defects to detect the direction of conventional WIMP dark matter; we also rely on existing WIMP techniques to identify the regions of interest within the crystal. Thus in our approach, the interrogation of crystal defects is restricted to a small number of events, as opposed to the continuous scanning necessary in  \cite{Tommer}. 


\end{document}